\documentclass[usenatbib,useAMSm a4paper, singlespacing]{mn2e}

\usepackage{caption}
\usepackage{subcaption}
 
\usepackage{graphicx}	
\usepackage{amsmath}	
\usepackage{amssymb}	
\usepackage{multirow,multicol}        
\usepackage{bm}		
\usepackage{pdflscape}
\usepackage[T1]{fontenc}
\usepackage{txfonts}

\usepackage{longtable,lscape, booktabs,etoolbox}
\usepackage{float}

\usepackage{rotating}
\usepackage{epsfig}
\usepackage{footnote}
\usepackage[referable]{threeparttablex}
\usepackage{hyperref} 
\usepackage{bookmark}

\def\mearth{\ifmmode {\rm M_{\oplus}}\else $\rm M_{\oplus}$\fi}
\def\Mearth{\ifmmode {\rm M_{\oplus}}\else $\rm M_{\oplus}$\fi}
\def\Rearth{\ifmmode {\rm R_{\oplus}}\else $\rm R_{\oplus}$\fi}
\def\Ms{\ifmmode {M_s}\else $M_s$\fi}
\def\Mp{\ifmmode {M_p}\else $M_p$\fi}
\def\Rp{\ifmmode {R_p}\else $R_p$\fi}
\def\rearth{\ifmmode {\rm R_{\oplus}}\else $\rm R_{\oplus}$\fi}

\def\aap{A\&A}

\def\apjl{ApJL}
\def\apjs{ApJS}

\newcommand{\Mdotacc}{\dot M_{acc}}

\newcommand{\Msun}{M_{\odot}}

\newcommand{\Lsun}{L_{\odot}}

\title[V883 Ori ]{The ALMA Early Science View of FUor/EXor objects. III. The Slow and Wide Outflow of V883 Ori \thanks{Based on ALMA observations, program number 2013.1.00710.S}}

\author[D. Ru\'iz-Rodr\'iguez et al.]
{ \Large D. Ru\'iz-Rodr\'iguez, $^1$\thanks{E-mail:dary.ruiz@anu.edu.au}
L. A. Cieza,$^{2,3}$
J. P. Williams,$^4$
D. Principe,$^{3,5}$
J. J. Tobin,$^{6,7}$ 
Z. Zhu,$^{8}$
A. Zurlo,$^{2,3,9}$
\\
$^{1}$Research School of Astronomy and Astrophysics, Australian National University, Canberra, ACT 2611, Australia\\
$^{2}$Millenium Nucleus ``Protoplanetary discs in ALMA Early Science", Chile \\
$^{3}$N\'ucleo de Astronom\'ia,  Facultad de Ingenier{\'i}a, Universidad Diego Portales,  Av. Ejercito 441, Santiago, Chile\\
$^{4}$Institute for Astronomy, University of Hawaii at Manoa, Honolulu, HI, 96822, USA\\
$^{5}$Department of Physics and Kavli Institute for Astrophysics and Space Research, Massachusetts Institute of Technology, Cambridge, MA 02139, USA\\
$^{6}$Homer L. Dodge Department of Physics and Astronomy, University of Oklahoma, 440 W. Brooks Street, Norman, OK 73019, USA\\
$^{7}$Leiden Observatory, Leiden University, P.O. Box 9513, 2300-RA Leiden, The Netherlands\\
$^{8}$Department of Physics and Astronomy, University of Nevada, Las Vegas, 4505 South Maryland Parkway, Las Vegas, NV 89154, USA\\
$^{9}$Universidad de Chile, Camino el Observatorio 1515, Santiago, Chile
}

\begin{document}

\date{}

\pagerange{\pageref{firstpage}--\pageref{lastpage}} \pubyear{2017}

\maketitle

\label{firstpage}

\begin{abstract}

We present Atacama Large Millimeter/ sub-millimeter Array (ALMA) observations of V883 Ori, an FU Ori object. We describe the molecular outflow and envelope of the system based on the $^{12}$CO and $^{13}$CO emissions, which together trace a bipolar molecular outflow. The C$^{18}$O emission traces the rotational motion of the circumstellar disk. From the $^{12}$CO blue-shifted emission, we estimate a \textit{wide} opening angle of $\sim$ 150$^{^{\circ}}$ for the outflow cavities. Also, we find that the outflow is very \textit{slow} (characteristic velocity of only 0.65 km~s$^{-1}$), which is unique for an FU Ori object. We calculate the kinematic properties of the outflow in the standard manner using the $^{12}$CO and $^{13}$CO emissions. In addition, we present a P Cygni profile observed in the high-resolution optical spectrum, evidence of a wind driven by the accretion and being the cause for the particular morphology of the outflows. We discuss the implications of our findings and the rise of these slow outflows during and/or after the formation of a rotationally supported disk.

\end{abstract}


\section{Introduction}

During the early stellar evolution process the key to understanding outflow motions is hidden. In stellar formation, these outflow motions might regulate the final stellar mass with a core-to-star efficiency of only 30$\%$ \citep{Offner2014}. In addition, it is believed that these outflows carry matter back to the molecular cloud, transporting energy and momentum to it, which may affect the dynamics of the surrounding envelope. However, the formation, evolution and effects of these flows is highy debated. Thus, a full understanding of the origin and evolution of these winds/outflows, might disentangle the unknown physical mechanisms that dictate the 1) low mass star formation efficiency in turbulent clouds \citep{Krumholz2012} and 2) an efficient transport of angular momentum to permit the accretion of matter onto the central star \citep{Blandford1982}. However, the physical origin(s) and features of these outflows are not well understood and our current knowledge of the entrainment process is limited due to the inability to trace the molecular gas a scale of a few au. In the ALMA era, observations of higher sensitivity and spatial resolution of young stellar objects surrounded by structures carved out by these outflows are required \citep[see][for a review]{Frank2014}. FU Orionis objects (FUors) are ideal candidates to observe and analyse due to their main characteristics: strong outflows and massive envelopes.

FUors are generally identified by their large and sudden increase of luminosity in optical light. This increase takes place in around $\sim$ 1$-$10 yrs and can amount to $\geq$ 5 mag in optical light. Although this optical variability has not been completely incorporated in the big picture of stellar formation and the evolution process, a large amount of matter ($\sim$ 0.01 M$_{\odot}$) accreting from the circumstellar disc onto the central object ($\sim$ 10$^{-4}$ M$_{\odot}$yr$^{-1}$), is the most likely cause of this variability  \citep{Hartmann1996}. These short events might be outbursts that are connected to the broad range of outflows observed in FUors. The surrounding envelope directly interacts with these outflows, which are likely the main dispersing mechanism of the natal circumstellar gas and dust, driving the evolution from a Class 0/I object to a Class II. From an observational perspective, an evolutionary trend in the opening angle of general protostellar outflows has been detected \citep{Velusamy1998, Arce2006, Seale2008}, where the outflow erodes the envelope and the widening of the cavity increases as the outflow ram pressure highly dominates over the infall ram pressure \citep{Arce2004}. The concept that cavities widen with time, postulates that Class 0 objects present opening angles ranging from 20$^{^{\circ}}$- 50$^{^{\circ}}$, Class I between 80$^{^{\circ}}$ and 125$^{^{\circ}}$ and Class II objects present outflows with cavities  $\geq$125 $^{^{\circ}}$ \citep{Arce2006}. Highly collimated and wide-angle molecular outflows differ in their gas velocities and mass. The former usually presents velocities on the order of $\rm v$$\sim$ 100$-$1000 km~s$^{-1}$, while the latter, less collimated outflows, are more massive with velocities on the order of $\rm v$ $\sim$10$-$30 km~s$^{-1}$ \citep[see][for a review]{Audard2014}. Theoretically, the observed widening in outflows might be connected to the interaction of highly accreting disc inner edges with a strongly magnetised central star, raising energetic winds. Among these models are the X winds  \citep{Shu2000}, disk winds \citep{Pelletier1992,Pudritz2007}, and accretion-powered stellar winds \citep{Ramanova2005}. More collimated outflows might be explained by a jet-driven bow shock, which essentially is an expanding bow shock produced by a dense and collimated jet that propagates through the ambient material, forming a thin shell of gas entrained in the wake of the outflow and extending from the jet head back to the star \citep{Raga1993, Ostriker2001}.

In addition, the detection of P Cygni profiles\footnote{Line profile composed of a red-shifted emission peak together with a blue-shifted absorption feature.} mainly in H$\rm \alpha$ and Na D lines are suggestive of energetic mass outflows/winds. These profiles, which are usually prominent in the spectra of FUor type stars \citep[e.g.][]{Calvet1993, Reipurth2002, Aspin2011}, are predicted by the presence of strong winds rising from the inner region of the disc \citep{Herbig1977, Bastian1985, Welty1992}. Therefore, the association of outflows and disc through energetic winds has begun one of the most promising scenarios to explain kinematic and dynamic motions at early stages of stellar formation.

\begin{table}
 \centering
 \begin{minipage}{62mm}
  \caption{V883 Ori Properties}
 \label{table:properties}
  \begin{tabular}{lcc}
\hline \hline \\[-3ex]
    \multicolumn{1}{c}{\textbf{Property}} &    
    \multicolumn{1}{c}{\textbf{Value}} &
     \multicolumn{1}{c}{\textbf{Reference}} \\[-0.2ex] 
     \multicolumn{1}{c}{\textbf{}} &
     \multicolumn{1}{c}{\textbf{}} &
      \multicolumn{1}{c}{\textbf{}}   \\[-2.4ex]\hline \hline 
R.A. (J2000)&05$^{h}$ 38$^{s}$ 18.10$^{s}$&1\\
Dec. (J2000) & -07$^{o}$ 02$^{'}$ ${26.00}''$&1 \\
$M_{*}$ ($M_{\odot}$) &1.3 $\pm$ 0.1&2 \\
$\rm \Mdotacc$  ($\rm \Msun yr^{-1}$)& 7.5e-5 & 2 \\
$\rm L_{*}$ ($L_{\odot}$)& 6&2\\  
$\rm L_{bol}$ ($L_{\odot}$)& 400&3\\  
$A_{V}$(mag) &19&4\\

 \hline 
 \end{tabular}
Reference: (1) 2MASS All-Sky Point Source Catalog,  (2) \citet{Cieza2016a}, (3) \citet{Strom1993}, (4) Spectral parameters from \citet{Caratti2012}.\\
\end{minipage}
\end{table}

As FUors are promising ``laboratories'' to contribute in the understanding of the envelope dissipation and core-to-star formation efficiency, we have conducted a new millimeter study of FUors and FUor-like stars presented in a series of papers by \citet{Zurlo2016}, \citet{RuizRodriguez2016}, \citet{Principe2016} and \citet{Cieza2016}. Here, we present ALMA band-6 (230 GHz/1.3 mm) continuum and  $^{12}$CO( J=2-1 ), $^{13}$CO( J=2-1 ) and  C$^{18}$O( J=2-1 ) line observations of an FUor type object identified initially as a faint star in the H$ \rm \alpha$ emission line survey of \citet{Haro1953} and designated as \textbf{V883 Ori}. We also report the optical spectrum of the $\rm H \alpha$ line at 6563 $\rm \AA$ taken with the MIKE  spectrograph \citep{Bernstein2003}. 

Since its detection, V883 Ori, located at a distance of 414 $\pm$7 pc \citep{Menten2007}, has been a source of major findings, thus providing hints about the formation and evolution of pre-main sequence stars. At first, its associated reflection nebulosity, IC 340, presented a morphological structure that suggested a star formation event involving the faint star, V883 Ori \citep{Haro1953}. Some years later, \citet{Nakajima1986} noticed a decrease in luminosity since V883 Ori was observed by \citet{Allen1975}. However, the first team to  describe this event and suggest this source as an FUor type object with a bolometric luminosity of $\sim$400 $L_{\odot}$ was \citet{Strom1993}. Although it was classified as an FUor type, no jet or molecular outflow was previously detected from V883 Ori \citep[e.g.][]{Sandell2001}, until these observations. More recently, \citet{Furlan2016} fitted the spectral energy distribution of V883 Ori (a.k.a HOPS-376) and they classified it as a flat spectrum protostar, where the mass of the envelope within 2500 au was found to be  2.87$\times$10$^{-2}$ $\rm \Msun$ and a cavity opening angle of $\sim$ 41$^{^{\circ}}$. As a part of the Protostellar Optical-Infrared Spectral Survey On NTT  (POISSON) performed by \citet{Caratti2012}, V883 Ori was included and using Br$\gamma$ as an accretion tracer, an equivalent width ($\rm EW$) of -3.6 $\rm \AA$ was found, corresponding to an accretion luminosity L$_{\rm acc}$(Br$\gamma)$ of 61.3 $\rm \Lsun$. More recently, \citet{Cieza2016a} described V883 Ori as a pre-main sequence object with a dynamical stellar mass of 1.3 $\pm$ 0.1 $M_{\odot}$ and photospheric luminosity of just $\sim$6 $L_{\odot}$ (based on the stellar mass, an assumed age of 0.5 Myr and the evolutionary tracks by \citet{Siess2000}).  Based on the stellar mass and the bolometric luminosity of 400 $L_{\odot}$, they derived an accretion rate of 7$\times10^{-5} \rm M_{\odot}$ year$^{-1}$, which is typical of FUor objects. More significantly, they reported the detection of the water snow line\footnote{Region of the disk where the temperature falls below the sublimation point of water.} at a distance of $\sim$ 42 au from the central star, a distance $\sim$10 times larger than expected for a disk passively heated by the stellar photosphere.

This relevant finding in an FUor ratified the importance of studying the evolution of circumstellar discs parallel to the outflows characteristic of these objects, in order to understand the main mechanisms involving the accretion flow and the high mass-loss rates. Table \ref{table:properties} summarises the estimated stellar parameters of V883 Ori.

We use $^{12}$CO and $^{13}$CO to describe the bipolar outflows of V883 Ori and the envelope material surrounding this source.  We present the results of these observations in this article organised as follows. Section \ref{Sec:Observations} describes the ALMA and MIKE observations, together with the reduction process. In Section \ref{Sec:Results}, we report the results obtained from interferometry and spectroscopy, additionally, we described the detected spectral features of this FUor. The implications and impact of our findings are discussed in Section \ref{Sec:Discussion}. The summary and conclusion are presented in Section  \ref{Sec:Summary}.

\begin{figure*}
\centering
\includegraphics[width=0.73\textwidth]{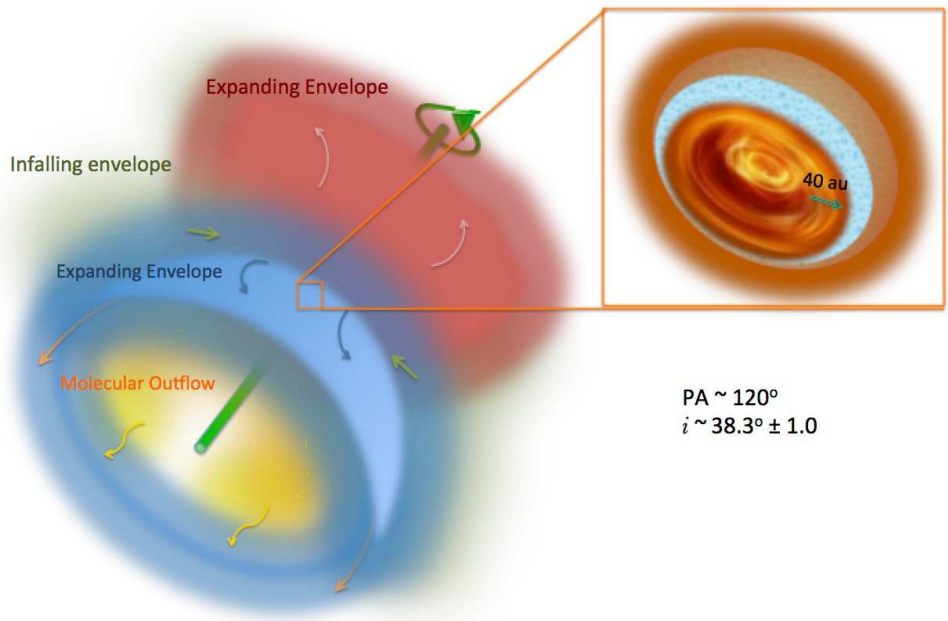}
\caption{Illustration showing the different dynamical and flux components traced by $^{12}$CO, $^{13}$CO and C$^{18}$O of V883 Ori. The outflows are coloured with red to illustrate the red-shifted emission, while blue illustrates the blue shifted emission. Envelope material close to and accreting onto the disc is coloured with green and its infalling motion is indicated by the small green arrows. The green line with a position angle of $\sim$ 120$^{^{\circ}}$ depicts the rotation axis of the entire system.  The inset shows a Keplerian disk probed by the C$^{18}$O emission, where a water snow-line at $\sim$40 au reported by \citet{Cieza2016a} is represented by the brown-blue gradient colour. Outward of the snow line, grain growth is accelerated by the high coagulation efficiency of ice-covered grains.}
\label{Fig:Cartoon}
\end{figure*}

\section[]{Observations}
\label{Sec:Observations}

\begin{figure*}
    \centering
    \hfill
    \begin{subfigure}[b]{0.495\textwidth}
        \centering
                 \includegraphics[width=1.0\textwidth]{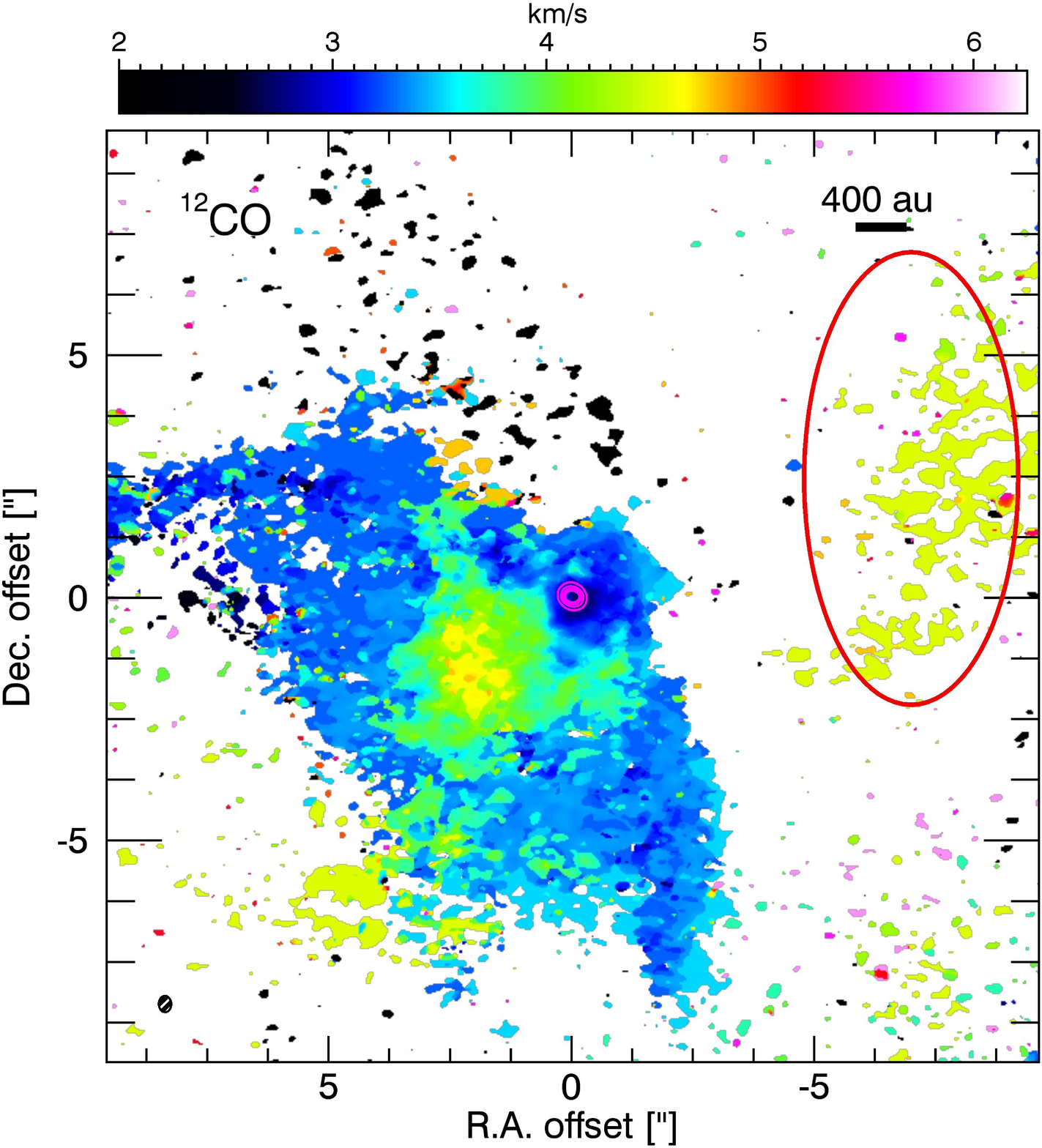}
                 \caption{ $^{12}$CO}
        \label{Fig:CO12}
    \end{subfigure}
    \hfill
    \begin{subfigure}[b]{0.495\textwidth}
        \centering
               \includegraphics[width=1.0\textwidth]{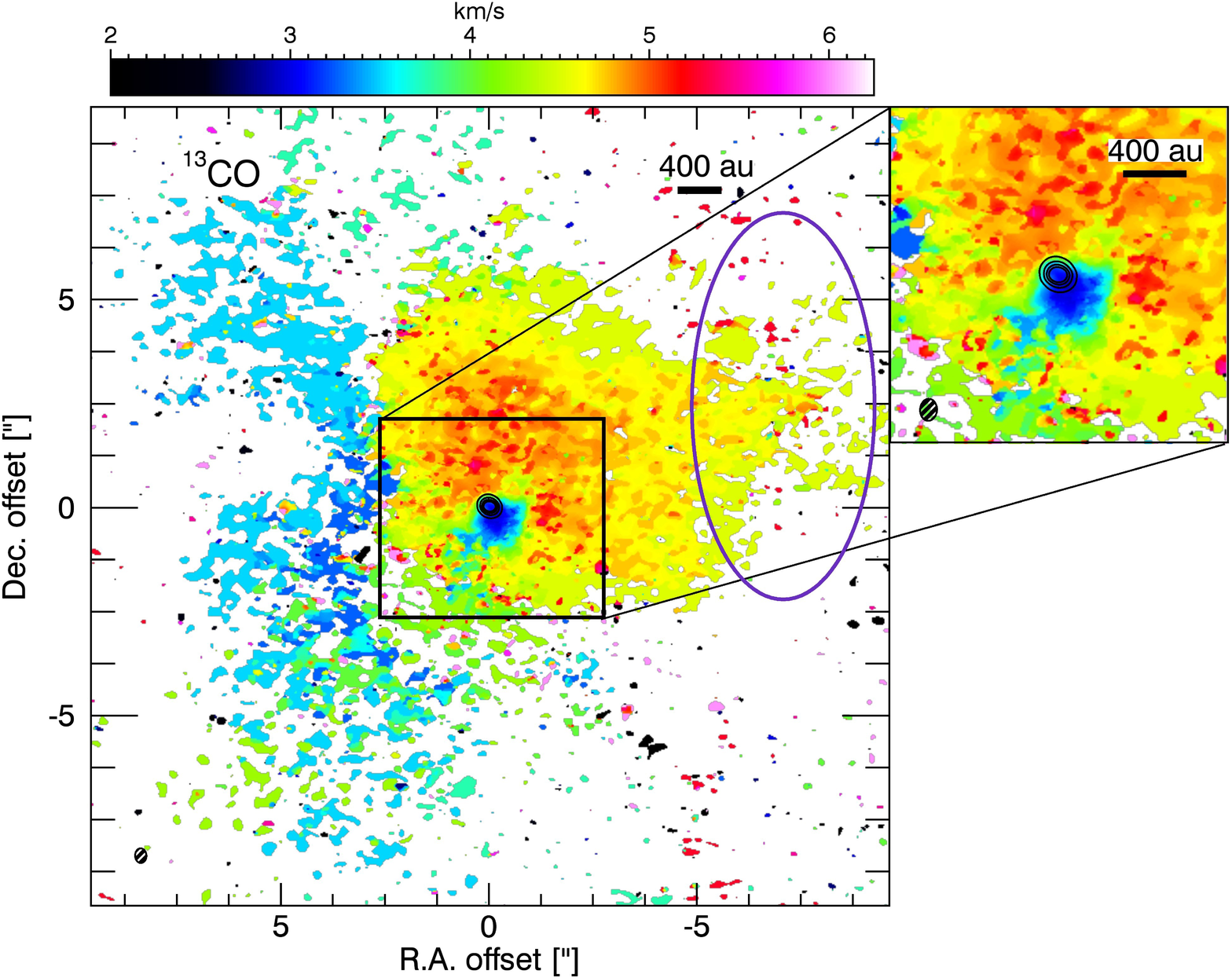}
                \caption{ $^{13}$CO }
        \label{Fig:CO13}
    \end{subfigure} 
    \hfill
       \begin{subfigure}[b]{0.495\textwidth}
        \centering
               \includegraphics[width=1.0\textwidth]{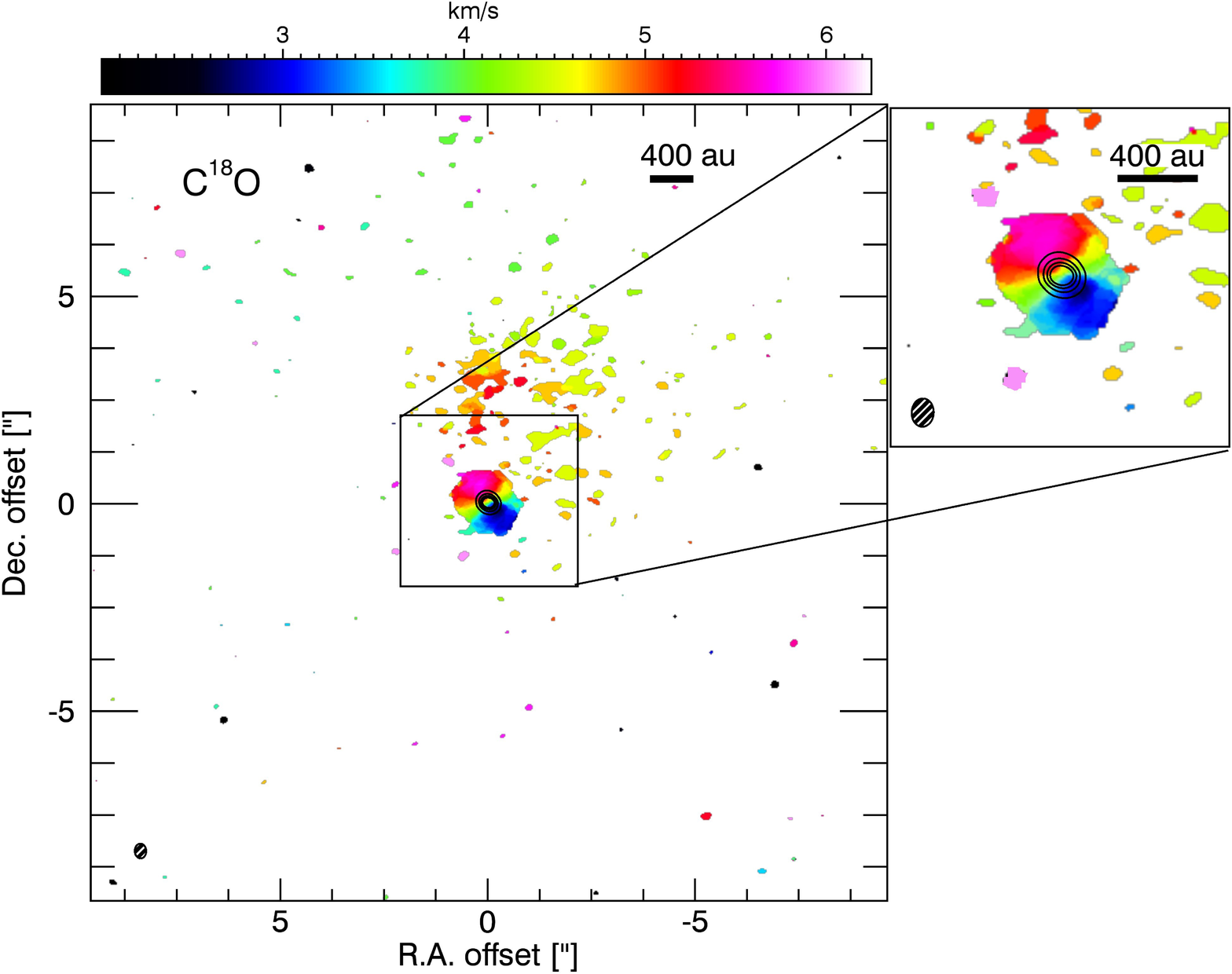}
                \caption{ C$^{18}$O }
        \label{Fig:CO18}
    \end{subfigure} 
       \caption{Figure a: $^{12}$CO velocity field map (moment-1) that was obtained from the integration over the velocity range from 2.0 to 6.25 km~s$^{-1}$. Figure b: Moment 1 of the $^{13}$CO emission integrated on the velocity range between 2.0 and 6.25 km~s$^{-1}$. Figure c: C$^{18}$O velocity field map integrated over the velocity range from 2.0 to 6.25 km~s$^{-1}$. The $^{12}$CO traces mostly the southern outflow with well defined edges, while the $^{13}$CO traces both the norther and southern outflows, but with a less defined shape. 
The C$^{18}$O emission reveals the Keplerian disk embedded within the envelope. Black contours show the continuum emission around V883 Ori at 10, 30, 80, 150 and 250 $\times$ rms (0.25 mJy beam$^{-1}$). The ${0.35}''\times{0.27}''$with P.A. $=$ -90$^{^{\circ}}$ synthesised beam is shown on the lower left corner of each panel. The upper right insets are a closeup of $\pm$ {2.7}'' for $^{13}$CO and $\pm$  {2.1}'' for C$^{18}$O of the central object. The purple and red ovals indicate the emissions traced by $^{12}$CO and $^{13}$CO as described in Sections \ref{Sec:co12results} and \ref{Sec:co13results}.} 
    \label{fig:three graphs}
  \label{Fig:CO}
\end{figure*}

\subsection{$^{12}$CO( J = 2-1 ), $^{13}$CO( J = 2-1 ) and C$^{18}$O( J = 2-1 ) Lines}

ALMA observations of V883 Ori, located at 05$^{h}$ 38$^{s}$ 18.10$^{s}$ -07$^{o}$ 02$^{'}$ ${26.00}''$, were taken under program 2013.1.00710.S during Cycle-2 over the course of three different nights. This program involves the observations of eight FUor/EXor objects:  V883 Ori \citep{Cieza2016a}, V2775 Ori \citep{Zurlo2016}, HBC 494 \citep{RuizRodriguez2016}, V1647 Ori \citep{Principe2016}, V1118 Ori, NY Ori,  V1143 Ori and ASASSN-13db \citep{Cieza2016}. Two of three observations were performed on December 12$^{th}$, 2014 and April 15$^{th}$, 2015 using 37 and 39 antennas on the C34-2/1 and C34/2 configurations, respectively. These configurations are similar with the shortest baseline of $\sim$ 14 m and longest of $\sim$ 350 m. The precipitable water vapor ranged from 0.7 to 1.7 mm with an integration time of $\sim$ 2 min per each epoch.

Additionally, a third observation of V883 Ori was performed on August 30$^{th}$, 2015 with 35 antennas in the C34-7/6 configuration with baselines ranging from 42 m to 1.5 km, an integration time of $\sim$3 min, and a precipitation water vapor of 1.2 mm. The quasars J0541-0541, J0532-0307 and/or J0529-0519 (nearby in the sky) were observed as phase calibrators.  J0423-013 and Ganymede were used as Flux calibrators, while the quasars J0607-0834  and J0538-4405  where observed for bandpass calibration. 

Our correlator setup included the J=2-1 transitions of $^{12}$CO, $^{13}$CO and C$^{18}$O centered at 230.5380, 220.3987, and 219.5603 GHz, respectively. The correlator was configured to provide a spectral resolution of 0.04 km~s$^{-1}$ for $^{12}$CO and of 0.08 km~s$^{-1}$ for  $^{13}$CO and C$^{18}$O.  The  total bandwidth available for continuum observations was 3.9 GHz.  The observations from all three nights were concatenated and processed together to increase the signal to noise and $uv$-coverage. The visibility data were edited, calibrated and imaged in CASA v4.4 \citep{McMullin2007}. The flux density calibration uncertainty is 10 $\%$.  We used the CLEAN algorithm to image the data and, using a robust parameter equal to zero, a briggs weighting was performed to adjust balance between resolution and sensitivity.  From the CLEAN process, we obtained the following synthesized beams:  ${0.35}''\times{0.27}''$with P.A. $=$ -90$^{^{\circ}}$ for $^{12}$CO, ${0.37}''\times{0.28}''$ with P.A. $=$ 86.5$^{^{\circ}}$ for $^{13}$CO and ${0.37}''\times{0.29}''$ with P.A. $=$ 87$^{^{\circ}}$ for C$^{18}$O. The rms is 12.5 mJy beam$^{-1}$ for $^{12}$CO, 16.0 mJy beam$^{-1}$ for $^{13}$CO and 13.9 mJy beam$^{-1}$ for C$^{18}$O. For the integrated continuum, we obtained a synthesized beam and rms of ${0.25}''\times{0.17}''$with P.A. $=$ -85.5$^{^{\circ}}$ and 0.25 mJy beam$^{-1}$, respectively.

\subsection{Optical Spectrum}

Additionally, we observed V883 Ori on the night of 29 February, 2016 with the Magellan Inamori Kyocera Echelle \citep[MIKE,][]{Bernstein2003}, a double echelle spectrograph at the Magellan (Clay) 6.5 m telescope, located in Las Campanas, Chile. This high resolution spectrograph covers a full optical wavelength range in the blue (320-480 nm) and the red (440-1000 nm) regime with spectral resolutions of 25000 and 19000, respectively. Our observations were taken with a slit size of 0.7$\times$5 $^{''}$ and the data have been binned 2$\times$2 in a slow readout mode with an exposure time of 1860 s. During the observation run were taken: Milky flats, Quartz flats, Twilight flats, ThAr comparison lamps and bias frames to use in the data reduction process. Thus, the data were bias-subtracted and flat-fielded to correct pixel to pixel variations by using the Carnegie Python tools\footnote{http://code.obs.carnegiescience.edu/mike} \citep[CarPy;][]{Kelson2003}. 

\begin{figure}
\centering
\includegraphics[width=0.48\textwidth]{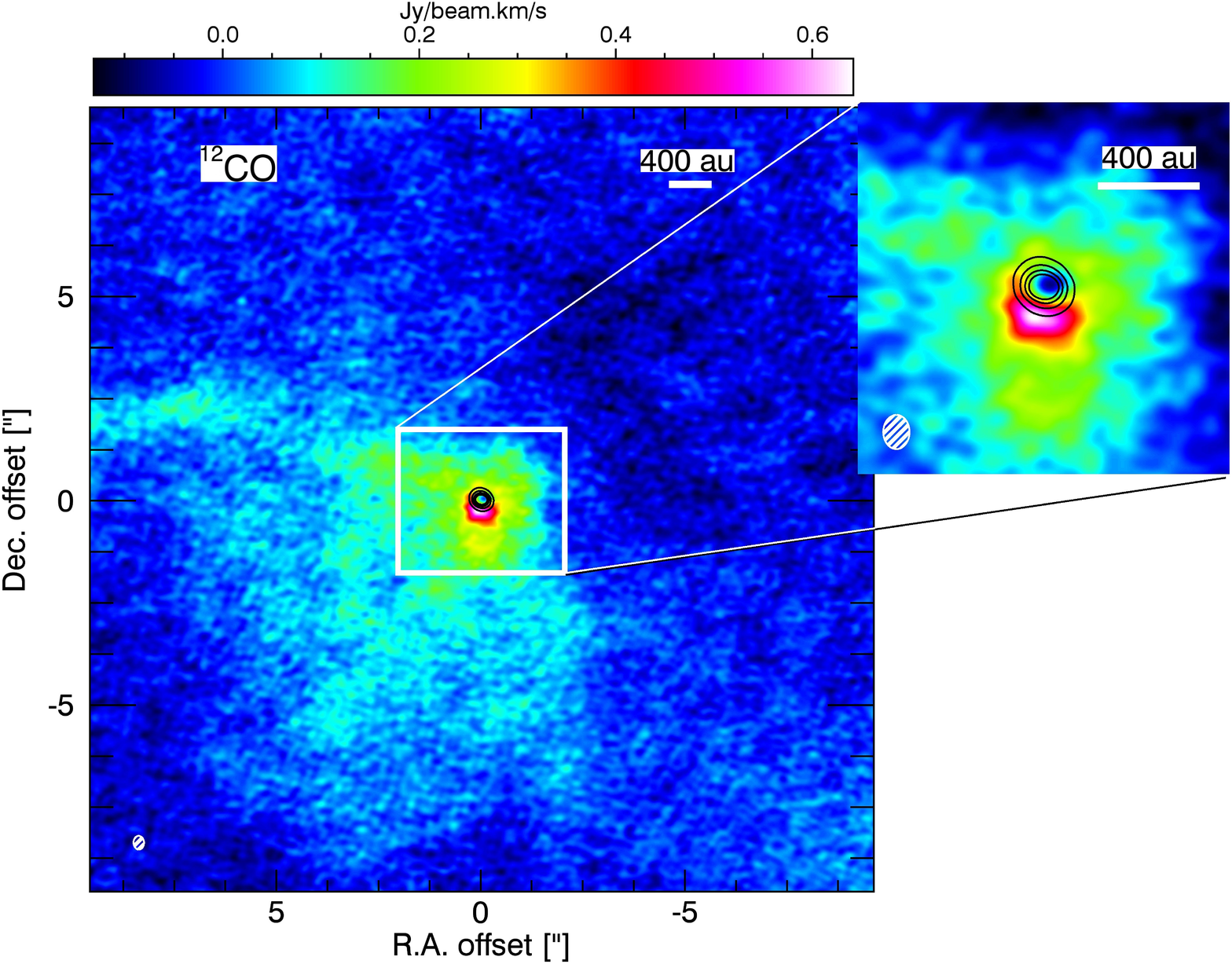}
\caption{$^{12}$CO  intensity maps (moment-0) integrated over the velocity range of 0.75 $-$ 8.0 km~s$^{-1}$. Black contours show the continuum emission around V883 Ori at 10, 30, 80, 150 and 250 $\times$ rms (0.25 mJy beam$^{-1}$). The upper right inset is a closeup ($\pm$ {1.8}'') of the central object. The ${0.35}''\times{0.27}''$ with P.A. $=$ -90$^{^{\circ}}$ synthesised beam is shown on the lower left corner.}
\label{Fig:CO12mom0}
\end{figure}

\section[]{Results}
\label{Sec:Results}

We obtained emission line profile data from V883 Ori of isotopologues $^{12}$CO, $^{13}$CO and C$^{18}$O with transitions $\rm J = 2 \rightarrow 1$ to trace the different components of this FUor object. Figure \ref{Fig:Cartoon} is a cartoon of the components detected with these optically thick and thin emissions, with a systemic velocity of 4.3 km~s$^{-1}$ \citep{Cieza2016a}. The CO emissions with bipolar shaped lobes, symmetrically placed around the central object (V883 Ori) are the product of the direct interaction between young outflows and the surrounding envelope, where the molecular outflows entrain part of the gas along the outflow axis independent of the physical origin. Then, from $^{12}$CO and $^{13}$CO emissions a bipolar shape is probed, where the cavity traced by the $^{12}$CO and pointing towards us, is less embedded in the surrounding envelope, while the cavity traced by the $^{13}$CO is more embedded than its counterpart. Unfortunately, we could not rule out faster gas towards the outflow axis from this data set or previous observations in databases. The colder and denser material close to the central object is traced by the $^{13}$CO and C$^{18}$O isotopes, where a disk with a Keplerian rotational profile is probed by the C$^{18}$O emission. For simplicity throughout the text, the blue- and red-shifted $^{12}$CO and $^{13}$CO emissions probing the bipolar cone are referred to as the southern and northern cavities, respectively. In order to estimate the outflow position angle (PA), we drew a line along the velocity gradient observed in C$^{18}$O and through the 1.3 mm continuum and thus, we obtained a PA of $\sim$ 120$^{^{\circ}}$ north through east \footnote{All position angles are specified north through east.}. From the 1.3 mm continuum emission, \citet{Cieza2016a} found a position angle of $\sim$ 32.4$^{^{\circ}}$ and from the major and minor axes of this emission an inclination ($i$) of $\sim$ 38.3$^{^{\circ}}$$\pm$ 1.0. Here, we adopted this inclination value to describe the orientation of the outflows.

\subsection{$^{12}$CO Emission}
\label{Sec:co12results}

\begin{figure*}
    \centering
     \begin{subfigure}{0.6\textwidth}
        \centering
        \includegraphics[width=1.0\textwidth]{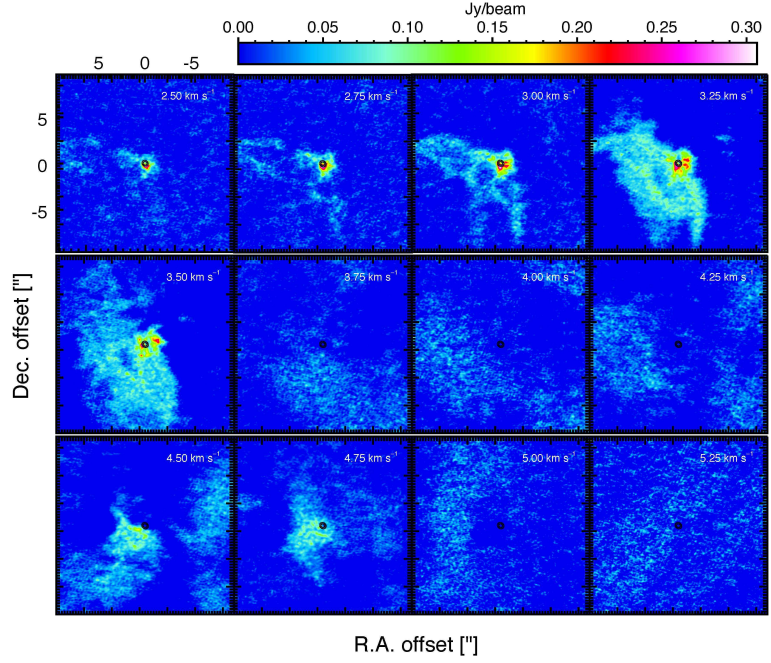}
        \caption{$^{12}$CO}
        \label{Fig:CanalCO12}
    \end{subfigure}
    \begin{subfigure}{0.6\textwidth}
        \centering
        \includegraphics[width=1.0\textwidth]{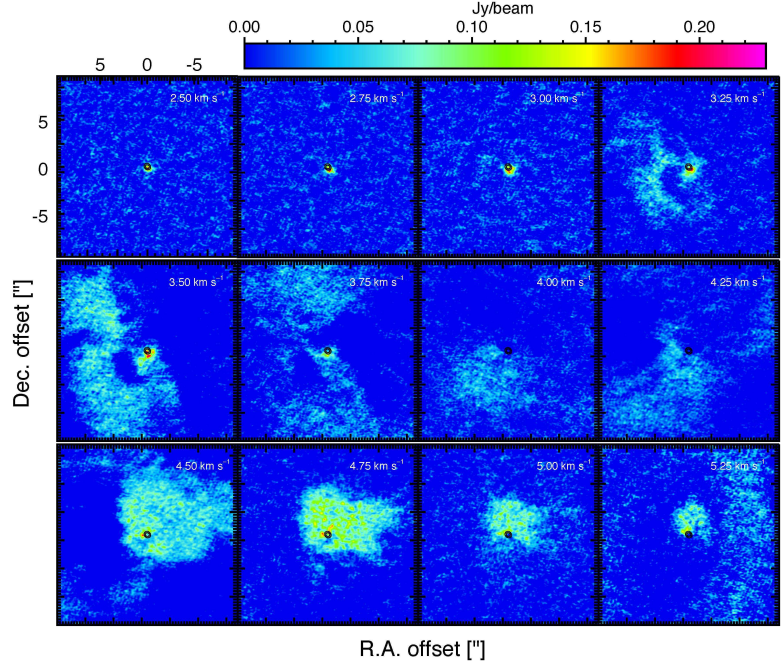}
        \caption{$^{13}$CO}
        \label{Fig:CanalCO13}
    \end{subfigure}
       \caption{Channel maps of the  $^{12}$CO and $^{13}$CO. LSR velocities are shown at the top-right corner of each panel with a systemic velocity of $\sim$4.3 km~s$^{-1}$. Black contours represent the  continuum emission around V883 Ori as shown in Figure \ref{Fig:CO12}.}
    \label{Fig:Canal}
\end{figure*}

Figure \ref{Fig:CO12} presents the integrated intensity of the weighted velocity (moment-1) maps for V883 Ori. This moment is integrated over the narrow velocity range with respect to the Local Standard of Rest (LSR) of 2.0 $-$ 6.25 km~s$^{-1}$, where the emission is detected at levels higher than 3$\sigma$ ($\sigma$ = 1.5$\times$10$^{-2}$ Jy beam$^{-1}$). This integration range for $^{12}$CO is chosen to match the moment-1 of $^{13}$CO and to display the kinematic structure of the $^{12}$CO and $^{13}$CO line emission, see Section \ref{Sec:co13results}. The $^{12}$CO emission tracing the southern molecular outflow has a  range between $\sim$ 0.75 km~s$^{-1}$ and $\sim$4.25 km~s$^{-1}$, while the red-shifted emission is observed in the range between 4.5 and 8.0 km~s$^{-1}$. The former emission is bright and extends up to the systemic velocity, which clearly traces the shape of the southern cavity. The latter is detected mostly close to the central object, and at what seems to be the end of the right arm of the outflow. However, emissions at velocities of $\sim$ 4.5 km~s$^{-1}$ are more likely to correspond to the parent molecular cloud. Thus, the emission indicated with a red oval in Figure \ref{Fig:CO12}, at a velocity of  $\sim$ 4.5 km~s$^{-1}$, seems to be better explained as being dominated by ambient emission rather than from an outflow emission, see also Figure \ref{Fig:Canal}. A slab of colder and denser envelope material located in the northern region of the outflow might be blocking the $^{12}$CO emission, making its interpretation ambiguous because the optically thin $^{13}$CO emission is brighter on this side of the object (see Section \ref{Sec:co13results} for more details). Another possibility is that the surrounding envelope might be built with different velocity components, and in the case of V883 Ori, the narrow velocity range of this emission causes the red-shifted $^{12}$CO emission on this side of the cavity to be spatially filtered out.

Figure \ref{Fig:CanalCO12} shows the $^{12}$CO channel maps, with a channel width of 0.25 km~s$^{-1}$, where the outflow cavity of a parabolic shape is more predominant at velocities of $\sim$ 3.25$-$3.50  km~s$^{-1}$. Interestingly, there is a slightly noticeable elongated feature towards the southeast, while in the $^{13}$CO channel maps a more pronounced feature is displayed in the same velocity range and location, which geometrically overlaps the $^{12}$CO feature, see Figure \ref{Fig:CanalCO13} . This feature seems to move further away from the central source, and could potentially be explained as outflowing layers entrained by a wide-angle wind. For this matter, we further explore kinematic features in the position-velocity (PV) diagrams to identify a possible parabolic PV structure, which can be produced by a wide-angle wind model at any inclination \citep{Shu2000, Lee2000}. Unfortunately, we did not find any signature of this characteristic, however, this does not rule out a radial wind producing a molecular outflow with a wide-opening angle.

From the southern cavity emission, we estimate an apparent opening angle of the outflow following the relation:

\begin{displaymath}
\theta \rm _{o} = 2 \tan ^{-1}\left [ \left ( 1 - e^{-1} \right )  \frac{R_{o}}{z_{o}}\right ]
\end{displaymath}

where $R_{o}$ is the transverse radius or radius to the rotation axis and $z_{o}$ is the distance along the outflow where the angle is measured  \citep[e.g.][]{Lee2005}. Thus, an opening angle of $\sim 150$$^{^{\circ}}$ is revealed beyond $\sim$ 600 au from the central source and an extension of 7300 au assuming a distance of 414 $\pm$7 pc \citep{Menten2007}. 

The narrow velocity range shown in the channel maps of $^{12}$CO data at velocity resolution of 0.25 km~s$^{-1}$, Figure \ref{Fig:CanalCO12}, leads to the conclusion that the outflow is not as highly energetic as other FU Ori objects in a similar evolutionary stage (Class I). 

\subsubsection{A hole in the $^{12}$CO emission?}
In Figure \ref{Fig:CO12mom0}, we present integrated intensity maps of the $^{12}$CO emission over the velocity range between 0.75 and 8.0 km~s$^{-1}$. Overall, this emission traces the southern outflow as described above,however, it also also presents a particular flux drop coincident with the central dust continuum peak. The inset of Figure \ref{Fig:CO12mom0} a zoomed in image of the hole, which is significantly weaker by a factor of $\sim$15 compared with the immediately surrounding emission. This feature is more likely due to dust absorption of the line emission and with a contrast of this magnitude, this implies that the dust continuum is considerably more optically thick than the CO emission around the central star.

\subsection{$^{13}$CO and C$^{18}$O Emissions}
\label{Sec:co13results}

Usually, millimeter observations of $^{12}$CO in star-forming environments tend to probe optically thick gas due to the high fractional abundance ($\rm \chi_{CO} \approx 10^{-4}$) of the isotope, while a less abundant CO isotope such as $^{13}$CO probes optically thin material. Thus, $^{13}$CO as a medium density tracer allows us to probe a higher density region than the low density tracer, $^{12}$CO. Figure \ref{Fig:CO13} shows the $^{13}$CO moment-1 map integrated in the velocity range between 2.0 and 6.25 km~s$^{-1}$. The $^{13}$CO gas has a bipolar distribution with respect to the outflow source on the northern side of V883 Ori, which covers the red-shifted velocities between 4.5 and 6.5 km~s$^{-1}$. This emission likely rises from the outer envelope that surrounds the protostellar core and the material interacting with the immediate surroundings of the outflow. The southern cavity (blue-shifted $^{13}$CO emission) is more diffuse and weaker than on the northern side. The southern cavity emits at velocities between 1.75 and 4.25 km~s$^{-1}$ and probes envelope material that might indicate that the outflow has been able to accelerate medium-density gas at large distances away from the central object, where the highest velocity components are observed close to the central object and ambient velocity components widen from the central source.

In figures \ref{Fig:CO12} and \ref{Fig:CO13} the moment-1 maps of $^{12}$CO and $^{13}$CO are shown, integrated over the same velocity range for comparison (2.0 $-$ 6.25 km~s$^{-1}$). The physical connection between the base of the cavity-envelope system traced by the $^{13}$CO isotope and what seems to be envelope material traced by $^{12}$CO, that reaches velocities of only 4.5 km~s$^{-1}$ , is indicated by a red oval in the $^{12}$CO moment-1 map and the physical location of the $^{12}$CO emission in the moment-1 of $^{13}$CO is indicated by a purple oval. The velocity field shows a gradual decrease in speed from the inside out.  At small radii, the outflow may be entraining inner envelope material, while at large distances from the central source it widens and acquires the systemic velocity.

\begin{figure}
\centering
\includegraphics[width=0.50\textwidth]{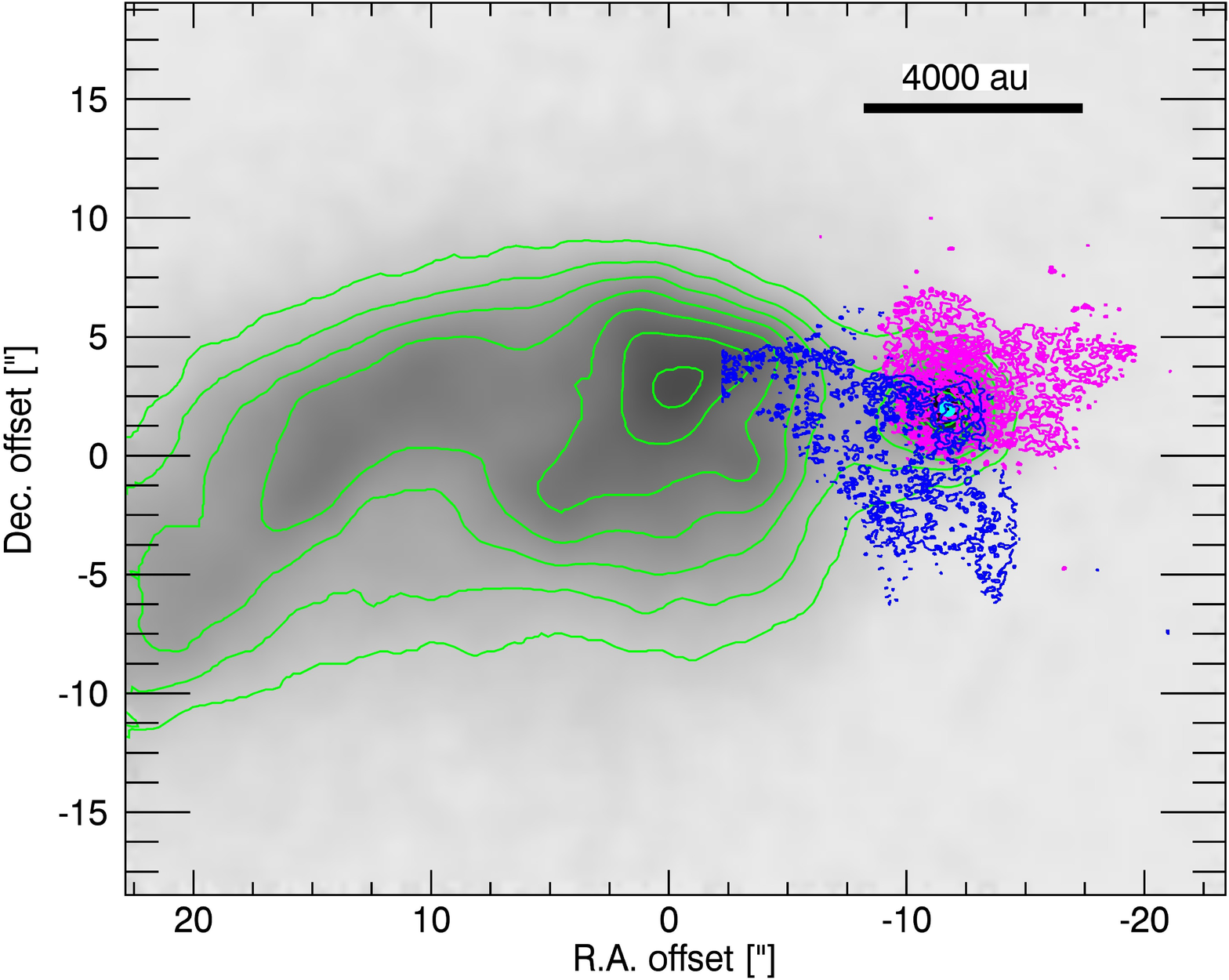}
\caption{Comparison of the 1.3 mm continuum, $^{12}$CO and $^{13}$CO emissions (contours) and the optical I-band \citep[0.75 $\micron$;][]
{Ahn2012} image of V883 Ori.  Blue and magenta contours show the integrated intensity of the $^{12}$CO and $^{13}$CO lines, respectively, at 20, 40, 80, 160, 240 $\times$ 3$\sigma$ levels. The cyan contours are the continuum emission and represent the position of V883 Ori. The green contours are the I-band data at 80, 150, 300, 450 and 600 $\times$ 3$\sigma$ levels.}
\label{Fig:SDSS9}
\end{figure}

Indeed, Figure \ref{Fig:Canal} presents the $^{12}$CO and $^{13}$CO channel maps, where the emission of both isotopes overlap and trace envelope-outflow material. Both, $^{12}$CO and $^{13}$ CO emissions peak on the central star, where the central disc is located. The blue shifted $^{12}$CO and $^{13}$CO line emission maps of V883 Ori reveal the low and medium density material expanding, part of which have already been noticed in optical images, since the northern outflow is not observed, such as in the I-band image shown in Figure \ref{Fig:SDSS9} where only the southern outflow appears illuminated.  An important implication of these emissions is that the cavity traced by the $^{12}$CO gas and pointing towards us, is less embedded in the surrounding envelope, while the cavity traced by the $^{13}$CO is more embedded than its counterpart. This is supported by the lack of detection of the $^{12}$CO gas and the bright detection of $^{13}$CO on the northern side of the object.

Besides the progressive dispersion of dense molecular gas at the northern region, an infalling and rotating motion is also observed close to the central star-disk system. It can be noted in the inset of Figure \ref{Fig:CO13} that at the base of the cavities, the $^{13}$CO gas probes a velocity gradient along the major axis of the 1.3 mm continuum, consistent with a Keplerian rotation and indicating the rotating and infalling material onto the central source. The infalling envelope near systemic velocity agrees with the rotating equatorial disc \citep{Cieza2016a}, also observed with C$^{18}$O, the lowest abundant isotope.

The densest material in V883 Ori is traced by the observations of the C$^{18}$O molecule, and presented in Figure \ref{Fig:CO18} as the moment-1 map integrated over the velocity range 2.0$-$6.25 km~s$^{-1}$. It is evident that this velocity map shows a structure delineating the material rotating around central object. In addition, the shape of the C$^{18}$O emission of these maps is very similar to the shape of the $^{13}$CO gas towards the base of the cavities.

\begin{figure}
\centering
\includegraphics[width=0.48\textwidth]{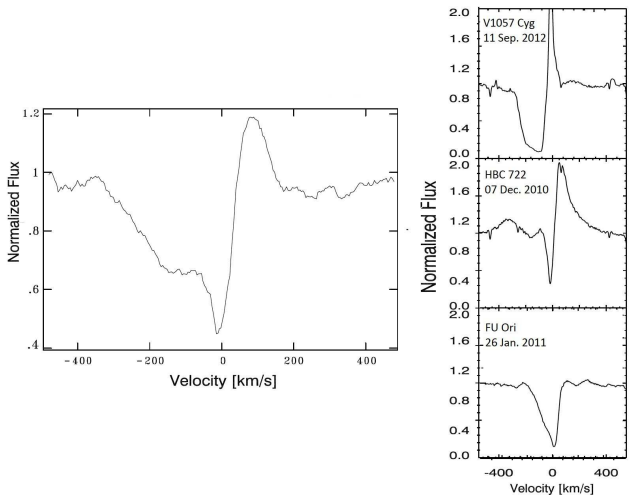}
\caption{V883 Ori spectrum of the $\rm H \alpha$ line at 6563 $\rm \AA$ taken with the MIKE spectrograph. On the right column,  we present the $\rm H \alpha$ velocity profiles of the FUor objects  V1057 Cyg, HBC 722,  and FU Ori itself. Spectra taken from \citet{Lee2015}.  }
\label{Fig:pcygni}
\end{figure}

\subsection{Optical Spectra}
\label{Sec:mike}

In order to study the winds detected in the optical regime, we obtained from MIKE the spectrum of one the most commonly observed outflow tracers, the $\rm H \alpha$ line at 6563 $\rm \AA$. Clearly visible in Figure \ref{Fig:pcygni} is the line characterised by a P-Cygni profile representing a wind/outflow, which is built of a strong and asymmetric blue-shifted absorption component together with a red-shifted component. The  slightly blue-shifted absorption feature shows a wing profile that extends to $\sim$$-$360 km~s$^{-1}$, while the emission line extends up to  $\sim$180 km~s$^{-1}$. The blue-shifted feature shows a very asymmetric line with the deepest absorption at $\sim$$-$14 km~s$^{-1}$ with the edge extending up to  $\sim$$-$65 km~s$^{-1}$, which remains relatively invariant until it reaches $\sim$$-$150 km~s$^{-1}$, then it weakens as the blue-shifted profile increases. Thus, the blue-shifted absorption seems to be composed of different features: a low-velocity and strong feature and a weaker structure fading as the velocity increases. The sharp change in the equivalent width ($\rm EW$) of this feature indicates a recent increase in mass loss rate of the system \citep[e.g.][]{Laakkonen2000}. On the other hand, the red-shifted emission peaks at $\sim$ 90 km~s$^{-1}$ and its intensity is not as strong as the blue-shifted absorption.

\subsection{Outflow Masses and  Kinematics}
\label{Sec:kinematics}

\begin{table*}
 \centering
 \begin{minipage}{169mm}
  \caption{Mass, Momentum, Luminosity and Kinetic Energy of the Outflow}
 \label{table:kinematics}
  \begin{tabular}{clccccccc}
 \hline \hline 
 \multicolumn{1}{c}{\textbf{}} &
  \multicolumn{1}{c}{\textbf{}} &
\multicolumn{2}{c}{\textbf{Blue shifted}\footnotemark[1]}    &
\multicolumn{2}{c}{\textbf{Red shifted}\footnotemark[2]}  &
\multicolumn{2}{c}{\textbf{Combined}} \\[0.5ex]

 \multicolumn{1}{c}{\textbf{Isotope}} &
 \multicolumn{1}{c}{\textbf{Property}} &
\multicolumn{1}{c}{\textbf{20 (K)} }  &
\multicolumn{1}{c}{\textbf{50 (K)}} &
\multicolumn{1}{c}{\textbf{20 (K)} }  &
\multicolumn{1}{c}{\textbf{50 (K)}} &
\multicolumn{1}{c}{\textbf{20 (K)}}    &
\multicolumn{1}{c}{\textbf{50 (K)}}\\[0.5ex]\hline \hline \\[-3ex]  

\multirow{5}{*}{\begin{sideways}{\Large{ $^{12}$CO}} \end{sideways}}
&Mass  (10$^{-2}$ M$_{\odot}$)  &             3.00 (172.20)\footnotemark[3]     & 4.50 (254.70)     &  1.00 (99.70) & 1.50 (147.50) &2.00 (136.00)&3.00 (201.10)\\
&Mass loss (10$^{-6}$ M$_{\odot}$ yr$^{-1}$)   &  1.15 (65.30)& 1.71 (96.70)  & 0.38 (37.80)& 0.56 (55.91)& 0.77 (51.60)& 1.14 (76.30)\\ 
&Momentum  ( 10$^{-2}$ M$_{\odot}$ km s$^{-1}$)   &  3.10 (170.02)&  4.55 (251.50) & 0.38 (32.00)  &  0.57 (47.40) & 1.74 (101.00) & 2.56 (149.50)  \\ 
&Energy (10$^{41}$ ergs.)   &                        3.18 (172.30)& 4.71 (255.00) & 0.32 (17.80)  & 0.48 (26.20) & 1.80 (95.10) & 2.60 (140.60)    \\
&Characteristic Velocity  (km s$^{-1}$)   &        1.03 (1.00) & 1.01 (1.00) & 0.38 (0.32)  & 0.38 (0.32) & 0.71 (0.66)& 0.70 (0.66)     \\
&Luminosity (10$^{-5}$ L$_{\odot}$)   &          9.95 (538.50) & 14.50 (797.00) & 1.00 (55.50) & 1.49 (82.10) & 5.50 (297.00) & 8.00 (439.50) \\[0.1ex]\hline

\multirow{5}{*}{\begin{sideways}{\Large{ $^{13}$CO}} \end{sideways}}
&Mass  (10$^{-2}$ M$_{\odot}$)  &  0.56 &0.85 & 3.65 & 5.53 & 2.22 & 3.20\\ 
&Mass loss ( 10$^{-6}$ M$_{\odot}$ yr$^{-1}$)  &  0.21  & 0.32  &1.39  & 2.10 & 0.80 & 1.21 \\ 
&Momentum  (10$^{-2}$ M$_{\odot}$ km s$^{-1}$) &    0.39  &0.59 &1.50 &2.23 & 0.94 & 1.41\\ 
&Energy (10$^{40}$ ergs.)   &    3.33  & 5.10   & 7.50 & 11.40& 5.42 & 8.30\\
&Characteristic Velocity  (km s$^{-1}$)   &        0.70 & 0.69 & 0.41  & 0.40 & 0.56 & 0.55     \\
&Luminosity (10$^{-5}$ L$_{\odot}$)   & 1.04&1.60  & 2.35&  3.56 & 1.70 & 2.58 \\[0.1ex]\hline

 \multirow{5}{*}{\begin{sideways}{\Large{ Total}} \end{sideways}}
&Mass  (10$^{-2}$ M$_{\odot}$)  &  3.56 &5.35 & 4.65 & 7.03& 4.22&6.20\\ 
&Mass loss ( 10$^{-6}$ M$_{\odot}$ yr$^{-1}$)  &  1.36   & 2.03   &1.77   & 2.66 & 1.57 & 2.35 \\ 
&Momentum  (10$^{-2}$ M$_{\odot}$ km s$^{-1}$) &   3.48 &5.13 &1.90&2.80 & 2.68 & 3.97 \\ 
&Energy (10$^{41}$ ergs.)   &   6.51 & 9.81   & 7.82 &11.88 & 7.22 & 10.90\\
&Characteristic Velocity  (km s$^{-1}$)   &        0.98 & 0.96 & 0.41  & 0.40 & 0.64 & 0.64     \\
&Luminosity (10$^{-5}$ L$_{\odot}$)   & 10.10&16.10  & 3.35&  5.05 &7.20 & 10.58 \\[0.1ex]\hline
\end{tabular}
$^{1}$ Blue-shifted outflow kinematics were estimated after a cut above 5$\sigma$ and integration of channels  between 1.5 and 4.25  km~s$^{-1}$ for $^{12}$CO and  $^{13}$CO.\\
$^{2}$ Red-shifted outflow kinematics were estimated with a threshold value above 5$\sigma$ and integration of channels between 4.5 and 7.0  km~s$^{-1}$ for $^{12}$CO and $^{13}$CO. \\
$^{3}$ Parameters inside the parentheses correspond to the computed values after applying the correction factors for optical depth effects to all the channels with emission above 5$\sigma$
\end{minipage}
\end{table*}

The fact that most of the $^{12}$CO emission in the southern cavity has a much higher intensity than the $^{13}$CO line, indicates that the $^{12}$CO line can be used as a tracer of the gas column density of the southern cavity. Likewise, the $^{13}$CO emission can trace the northern side of V883 Ori.  Considering that both the $^{12}$CO and $^{13}$CO lines trace the bipolar cavity, we use these emissions to derive estimates of the mass of the outflow, $\rm M_{flow}$, and its kinematic properties (kinetic energy, $\rm E_{flow}$, momentum, $\rm P_{flow}$, and luminosity, $\rm L_{flow}$) in the standard manner \citep[e.g.][]{Cabrit1990, Dunham2014}. Thus, following the process described in Section 3.4 in \citet{RuizRodriguez2016}, we estimate these quantities from the blue- and red-shifted emissions, separately. However, as is often stated, the $^{12}$CO emission is optically thick and to derive accurate gas column densities, it is necessary to correct for the optical depth of this line. For that matter, $^{13}$CO, as an optically thin tracer, is used to correct for optical depth effects in the $^{12}$CO data. Hence, after computing the ratio of the brightness temperatures:

\begin{displaymath}
 \frac{\textrm{T}_{12}}{\textrm{T}_{13}} = X_{12,13}\frac{1- \textrm{exp}(-\tau _{12})}{\tau _{12}}
\end{displaymath}

where the abundance ratio $X_{12,13}$ = [$^{12}$CO]/[$^{13}$CO] is taken as 62 \citep{Langer1993}, from all the channels with detection above 3.5$\sigma$. We also consider that $^{12}$CO and $^{13}$CO probe opposite regions in the bipolar shape of V883 Ori, meaning that the number of channels with a computed ratio is small because $^{12}$CO and $^{13}$CO trace different regions at a narrow velocity range. Therefore, in order to apply the correction factor to all the channels with $^{12}$CO detection, it is necessary to extrapolate values from a parabola fitted to the weighted mean values of the form

\begin{displaymath}
\centering
\frac{\textrm{T}_{12}}{\textrm{T}_{13}} = 0.57 + 0.34(\textrm{v-v}_{\textrm{\textsc{LSR}}})^{2}.
\label{eq:parabola}
\end{displaymath}

\begin{figure}
\centering
\includegraphics[width=0.5\textwidth]{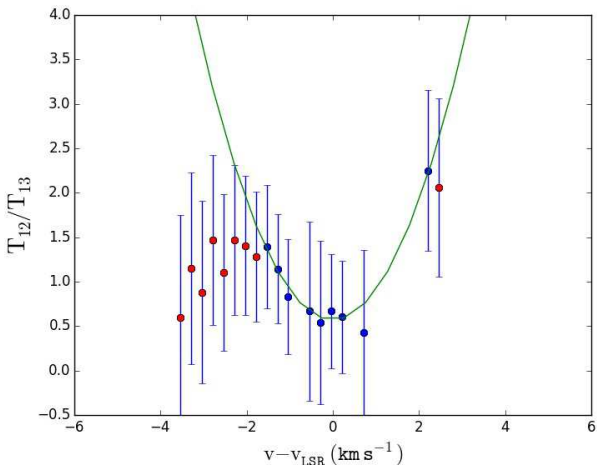}
\caption{Ratio of the brightness temperatures $\frac{T_{12}}{T_{13}}$ as a function of the velocity from the systemic velocity. The blue solid  dots are the weighted mean values and the error bars are the weighted standard deviations in each channel. The red solid dots are weighted mean values not used in the fitting process. The green solid line is the best-fit second-order polynomial with a $\chi^{2}$ of 0.6.}
\label{Fig:factor}
\end{figure}

In the fitting process, the minimum ratio value was fixed at zero velocity and we did not include those data points presented as the red dots in Figure \ref{Fig:factor}, because at these velocities $^{12}$CO starts becoming optically thin. The best fit with a $\chi^{2}$ of 0.6 is shown in Figure \ref{Fig:factor} as a solid green line, where the blue dots correspond to the weighted mean values and the error bars are the weighted standard deviations in each channel. In this particular case, the fitted parabola and the derived outflow parameters must be taken with caution because of the poor fitting, which highly depends on the weighting of the last data point to the right, see Figure \ref{Fig:factor}. This is because the $^{13}$CO emission is usually not detectable or is very weak in most mapping positions and velocities where the $^{12}$CO emission is detected, and vice versa.

To assure we are using the emission mostly from the outflow, we performed a first cut to values above 5$\sigma$ and the integration of channels between velocities ranging from 1.5 to 4.25  km~s$^{-1}$ and between 4.5 and 7.0 km~s$^{-1}$. In order to obtain a total estimate of these values, the range and number of channels in the integration are the same for $^{12}$CO and $^{13}$CO. The characteristic velocity of the outflow of $\sim$0.65 km~s$^{-1}$ is estimated using $\rm v_{flow}=\frac{P_{flow}}{M_{flow}}$, where $\rm P_{flow}$ and $\rm M_{flow}$ are the momentum and mass of the outflow, respectively \citep[e.g.][]{Dunham2010}. Taking the extent of the $^{12}$CO blue-shifted emission  of $\sim$7300 au (17.5${''}$) and the maximum speed of the gas extension, obtained using $\frac{\rm v_{13}- v_{12}}{2}$ where $\rm v_{13}$ and $\rm v_{12}$ are the $^{13}$CO red- and $^{12}$CO blue-shifted maximum velocities, we estimated a kinematic age for V883 Ori of $\sim$10000 years to obtain the mechanical luminosity and mass loss rate of the outflow. Here, it is important to note that the estimate of the dynamical timescale is a lower limit since we have only used our ALMA data and ignored the apparent extension observed in optical wavelengths of $\sim$ 64${''}$, see e.g. Figure \ref{Fig:SDSS9}. Thus, the outflow with an apparent extension of $\sim$ 27000 au, could be 4 times older than our estimate. We note that this estimated age  ($>$ 10$^{4}$ yr) is larger than the typical duration of an FU Ori outburst ($\sim$ 10$^{2}$ yr). This implies that the ongoing accretion outburst cannot be directly responsible for the properties of the observed outflow. Table \ref{table:kinematics} shows the estimates\footnote{Properties not corrected for inclination and optical effects.} at temperatures of 20 and 50 K. The actual values could be higher than those listed in Table \ref{table:kinematics}  because the estimated properties highly depend on the true values of the outflowing gas temperatures for both the $^{12}$CO and $^{13}$CO lines, and our observations have a maximum resolvable angular scale (MRS) of $\sim$11${''}$, meaning that a fraction of the total outflow emission might be missing. In other words, the outflow cavities are $\sim$15${''}$ across, and thus larger than the MRS. Hence, an extended component ($>$11${''}$) between the outflow cavities might not be resolved out. Future observations with the ALMA Compact Array would be useful to image the outflow at larger angular scales. In addition, taking into account that the difference between the outflow and envelope emission is marginal based on the small number of channels of $^{12}$CO  and $^{13}$CO with emissions above 3$\sigma$, these estimates could be contaminated by envelope emission.

\section[]{Discussion}
\label{Sec:Discussion}

\subsection{The extension and velocity of the outflow in the V883 Ori system}

From Figure \ref{Fig:CO}, it is evident that the southern cavity is traced by the $^{12}$CO emission, while the $^{13}$CO emission traces the shape of the northern cavity. Together, these emissions delineate the bipolar outflow of V883 Ori. In optical images \citep{Ahn2012}, the southern cavity seems to have a much larger extension than in our ALMA data. However, this could be an effect of the illumination caused by the interaction of the dusty material and the high luminosity of the central object. Figure \ref{Fig:SDSS9} shows a comparison of the optical image and the bipolar outflow ($^{12}$CO and $^{13}$CO emissions), and although the outflows roughly coincide in the projection over the I-band image, comparing these observations is not straightforward and thus, it does not ensure that the I-band image shows the real and physical extension of the cavity. Also, it is worth noting that the relatively limited field of view (FOV) of our ALMA data of only 0.1 arcminutes$^{2}$ cannot be compared well with other facilities with considerably larger FOV, such as the I band image shown in Figure \ref{Fig:SDSS9} with FOV of $\sim$ 0.5 arcminutes$^{2}$. Future mosaicking observations with ALMA or imaging by interferometers with a larger field-of-view, such as the Submillimeter Array, are needed to better determine the real extension of the outflow.

 From the $^{12}$CO blue-shifted emission, an opening angle of $\sim$150$^{^{\circ}}$ is estimated, where one of the most striking characteristics of the outflows is the relatively slow velocity with a characteristic velocity of only $\sim$ 0.65 km~s$^{-1}$, see Figure \ref{Fig:Canal}. The typical FUor outflow velocity ranges between 10 and 40 km~s$^{-1}$ with a wide range of collimation \citep{Evans1994, Zurlo2016, RuizRodriguez2016}, although, some FUors do not show CO emission associated with outflows (e.g. FU Orionis itself does not show an outflow \citep{Evans1994}). While our observational findings can be used as inputs to test slow-velocity outflows in FU Ori objects, yet, we are unable to compare to other FUors with similar outflow features because to date, these wide and slow-outflows have only been detected in V883 Ori. Thus far, these low-velocity outflows have been observed only in other Class 0/I objects, such as Per-Bolo 58 (2.9 km~s$^{-1}$), CB 17 MMS (2.4 km~s$^{-1}$), L1451-mm (1.3 km~s$^{-1}$), L1148-IRS (1.0 km~s$^{-1}$), L1014-IRS (1.7 km~s$^{-1}$) \citep[][and references therein]{Dunham2011}. However, these are younger, still embedded cores, with low luminosity and are not experiencing high accretion rates (i.e. outburst). V883 Ori is a stellar object of 1.3 $\rm \Msun$ and extremely luminous, accreting a large amount of matter onto the central source through a disk with Keplerian rotation \citep{Cieza2016a}. This Keplerian disk, from the 1.3 mm continuum map, is observed without asymmetries, discarding a stellar companion influencing the geometry and kinematics of the outflows. Thus, the physical mechanism responsible for these particular slow and wide-angle outflows must be triggered during and/or after the formation of a rotationally supported disk.

Recently, a similar opening angle was observed in the FU Ori Class I, HBC 494 \citep{RuizRodriguez2016}. The authors attributed the wide opening angle due to the presence of energetic winds as a result of the interaction of highly accreting disc inner edges with a strongly magnetised central star \citep{Snell1980, Blandford1982, Shu2000}. However, HBC 494 presents a very energetic outflow with a velocity gradient perpendicular to the outflow axis of rotation, while V883 Ori does not harbour an energetic driving source. This can be noted in the moment-1 of the $^{12}$CO and $^{13}$CO emissions shown in Figure \ref{Fig:CO} and obtained from an integration of a very narrow velocity range (2.0 - 6.5 km~s$^{-1}$). This narrow emission suggests that the triggering mechanism of these wide opening outflows in V883 Ori might have occurred 1) a long time ago, where another FUor outburst event could take place with an average time span of thousands of years between outbursts \citep{Scholz2013} or 2) in a quiescent disk without the creation of a high velocity outflow component, which might be related to rotation of the central protostar \citep{Ramanova2005, Konigl2011}. Unfortunately, there is not a record of the outflow onset or evidence of a high velocity component emission to rule out and/or confirm any of these possibilities.

While the narrow velocity of the $^{12}$CO and $^{13}$CO emission implies a slow outflow (see Figure \ref{Fig:Canal}), it also impacts the estimates of the mass and kinematic parameters  shown in Table \ref{table:kinematics} (i.e. mass-loss rate, the mechanical luminosity, momentum, and kinetic energy). Nevertheless, these values are on the order of other FU Ori objects such as, V2775 Ori, L1165, HBC 494 \citep{Zurlo2016, Dunham2014, RuizRodriguez2016}. Similarly, compared to previous studies a total outflow mass in the range of  10$^{-4}$ and 10$^{-1} \Msun$ is typical of outflows in other young stars \citep[e.g.][]{Wu2004, Curtis2010, Arce2006, Dunham2014, Klaassen2016}. Inspecting  Table \ref{table:kinematics}, the outflow parameters increased by a factor of $\sim$60-70, after correcting for optical depth effects, in agreement with previous results that established that these outflow parameters can increase by factors of up to 90 after correcting for inclination and optical effects \citep[e.g.][]{Curtis2010, Dunham2014, RuizRodriguez2016}. However, it is not easy to directly compare these parameters because 1) uncertainties in the method used and 2) the estimates in the literature differ by observing method, i.e. single dish vs. interferometer observations. For instance, parameters from single-dish data may take contributions from the extended cloud emission, increasing these estimates by a few factors when compared with the smaller scales sampled by the interferometer. Therefore, a more complete characterisation of the kinematics and dynamics of the outflows in FUors is required in the near future.

\subsection{Comparison with other P-Cygni profiles.}

In general, the presence of a P-Cygni optical profile indicates powerful winds likely rising from the disc \citep{Hartmann1996}, allowing the accretion of matter onto the central stellar core \citep[e.g.][]{Bai2013}.  As P-Cygni profiles have been observed in $\rm H \alpha$ lines of FUors such as FU Ori, V1057 Cyg and HBC 722 \citep{Herbig2003, Powell2012, Miller2011, Lee2015}, here we compare our spectrum to those objects as is shown in Figure \ref{Fig:pcygni}. These spectra were previously presented in \citet{Lee2015} and observed with the High Resolution Spectrograph \citep{Tull1998} of the Hobby$-$Eberly Telescope (HET) \citep{Ramsey1998} at McDonald Observatory and the Bohyunsan Optical Echelle Spectrograph (BOES) at Bohyunsan Optical Astronomy Observatory.

 Although comparing the $\rm H \alpha$ profile of these objects is difficult because they differ in: 1) time from last outburst and 2) amount of envelope material; these objects have shown observational evidence of the main features and variability of their profiles. For instance,  HBC 722 is an FU Ori object observed pre- and post-outburst \citep{Cohen1979, Semkov2010} and thus, it offers the opportunity to compare the spectrum of an FUor in a quiescent state and during the outburst ($\sim$ 6 years from outburst). The spectra of this object have changed significantly pre- and during outburst. In short, pre- and during outburst the $\rm H \alpha$ profile remained mostly in emission, while decreasing its $\rm EW$ \citep[between Aug and Sept. 2010;][]{Semkov2010}, until finally a few months later they acquired the shape of a P-Cygni profile \citep[on Dec. 2011;][]{Semkov2012}. Subsequently, the detected P-Cygni profile presented a constant strength variability in the following years, \citep[more details in][]{Lee2015}. In Figure  \ref{Fig:pcygni}, it can be noted that the blue-shifted absorption feature is considerably less broader than the $\rm H \alpha$ profile of V883 Ori. Then, if the $\rm EW$ strongly depends on the physical events taking place around the central star, an increase in the mass loss rate might broaden the $\rm EW$ \citep{Laakkonen2000}. That is the case of the $\rm H \alpha$ profile of V1057 Cyg, which also has varied in time since its outburst ($\sim$ 40 years from outburst)  \citep{Laakkonen2000}. The width of this profile is more similar to the $\rm H \alpha$ of V883 Ori, however, the latter shows a particular shape, see Figure \ref{Fig:pcygni}. This peaked feature seems to be built by different components and located at different distances from the central star, one is a weaker and high velocity component and the other(s) is a strong and low velocity component(s). In fact, it has been argued that a narrow central absorption comes from the central object and the ``wings'' correspond to the disc \citep{Lee2015}.  However, V883 Ori with a bolometric luminosity of 400 $\Lsun$ \citep{Strom1993} complicates the identification of these components, independently.

On the other hand, the $\rm H \alpha$ profile of FU Ori highly differs from the $\rm H \alpha$ profile of V883 Ori. To begin, a P-Cygni profile has vanished almost completely, where the red-shifted emission line has decreased considerably ($\sim$ 79 years from outburst). A similar feature was observed in HBC 722, soon after the outburst when the wind diminished, leaving mostly an $\rm H \alpha$ absorption profile \citep{Lee2015}. Although, we cannot directly compare or make a conclusion about the evolutionary state of the system, the $\rm H \alpha$ profile of V883 Ori indicates the presence of strong and persistent winds, which might be related to the wide-opening angle of the outflows. In fact, if the $\rm H \alpha$ absorption profile of V883 Ori arises solely via the accreting shock on the stellar photosphere, this would lead to the launching angle of the wind to be $\gtrsim$52$^{^{\circ}}$.

\subsubsection{P Cygni and Slow Winds.}

 In general, the blue-shifted absorption features in FUors highly depend on the velocity shift, where larger blue-shifts are related to the strength of the profile line \citep{Petrov1992, Calvet1993, HartmannCalvet1995}. If one assumes the strong winds originate in the accretion disk, the strongest lines show the largest expansion velocities, while the weak lines originate close to the disk photosphere. Potentially, magnetic fields anchored in the rotating disc itself could accelerate disk winds outwards  \citep{Blandford1982, Shu2000}. However, the slow winds in V883 Ori might originate in the outer part of the disk, where the location of the footpoints of wind-launching magnetic field lines on the disk, might determine the velocity components of the system.

\section[]{Summary}
\label{Sec:Summary}

In this paper, we have presented the results of the ALMA observations, together with the optical spectrum of V883 Ori. This object is an FU Ori source with a \textit{wide} opening angle of $\sim$ 150$^{^{\circ}}$ (measured east through north) with an extension of $\sim$7300 au that was detected from the $^{12}$CO blue-shifted emission. From the $^{12}$CO and $^{13}$CO emissions a bipolar shape of the outflow cavities is traced, while C$^{18}$O emission probes a Keplerian circumstellar disk. V883 Ori is a unique FU Ori object because it presents such a slow outflow with a characteristic velocity of only 0.64 km~s$^{-1}$. This is surprising as current models predict outflow velocities of around 10$-$50 km~s$^{-1}$ \citep[e.g.][]{Pudritz1986, Federrath2014}. Therefore, further theoretical and  observational studies are needed to investigate the origin of the slow and wide angle outflow in V883 Ori. A P Cygni profile observed in the $\rm H \alpha$ line centred at 6563 $\rm \AA$ provides evidence of the presence of winds likely rising from the disc and being the physical mechanism responsible for the morphology of the outflows. We estimate the kinematic properties of the outflow in the standard manner, these values are on the order of other FUors and young stars with outflows; after these parameters were corrected for optical effects, they increased by a factor of $\sim$ 60-70. However, as discussed in Section \ref{Sec:kinematics}, this optical depth correction must be taken with caution.

\section*{Acknowledgments}

We are grateful to an anonymous referee for suggestions that helped improve the quality of this article. L.A.C.,  D.P., and A.Z., acknowledge support from the Millennium Science Initiative (Chilean Ministry of Economy),  through grant Nucleus  RC130007. L.A.C. was also supported by CONICYT-FONDECYT grant number 1140109. D.P. was also supported by FONDECYT grant number 3150550. J.J.T acknowledges support from the University of Oklahoma, the Homer L. Dodge endowed chair, and grant 639.041.439 from the Netherlands Organisation for Scientific Research (NWO).

This paper makes use of the following ALMA data:
ADS/JAO.ALMA No. 2013.1.00710.S . ALMA is a partnership of
ESO (representing its member states), NSF (USA) and NINS (Japan),
together with NRC (Canada), NSC and ASIAA (Taiwan), and
KASI (Republic of Korea), in cooperation with the Republic of
Chile. The Joint ALMA Observatory is operated by ESO, AUI/NRAO
and NAOJ. The National Radio Astronomy Observatory is a facility
of the National Science Foundation operated under cooperative
agreement by Associated Universities, Inc.

Funding for the Sloan Digital Sky Survey IV has been provided by the
Alfred P. Sloan Foundation, the U.S. Department of Energy Office of
Science, and the Participating Institutions. SDSS acknowledges
support and resources from the Center for High-Performance Computing at
the University of Utah. The SDSS web site is www.sdss.org.
SDSS-III is managed by the Astrophysical Research Consortium for the Participating Institutions of the SDSS-III Collaboration including the University of Arizona, the Brazilian Participation Group, Brookhaven National Laboratory, Carnegie Mellon University, University of Florida, the French Participation Group, the German Participation Group, Harvard University, the Instituto de Astrofisica de Canarias, the Michigan State/Notre Dame/JINA Participation Group, Johns Hopkins University, Lawrence Berkeley National Laboratory, Max Planck Institute for Astrophysics, Max Planck Institute for Extraterrestrial Physics, New Mexico State University, New York University, Ohio State University, Pennsylvania State University, University of Portsmouth, Princeton University, the Spanish Participation Group, University of Tokyo, University of Utah, Vanderbilt University, University of Virginia, University of Washington, and Yale University. 

\bibliographystyle{mn2e} 
\bibliography{biblio}

\bsp

\label{lastpage}

\end{document}